# Formation of aromatics in rich methane flames doped by unsaturated compounds

H.A. Gueniche, P.A. Glaude*, R. Fournet and F. Battin-Leclerc

Département de Chimie-Physique des Réactions, UMR n° 7630 CNRS, INPL-ENSIC
1, rue Grandville - BP 20451 - 54001 NANCY Cedex – France

## Abstract

In order to better understand the importance of the different channels leading to the first aromatic ring, we have investigated, the structure of a laminar rich premixed methane flame doped with several unsaturated hydrocarbons: allene and propyne, as they are precursors of propargyl radicals, which are well known as having an important role in forming benzene, 1,3-butadiene, to put in evidence a possible production of benzene due to reactions of $C_4$ compounds, and, finally, cyclopentene, which is a source of cyclopentadienyl methylene radicals which are supposed to easily isomerizes to give benzene. A ratio additive / $CH_4$ of 16 % and an equivalence ratio of 1.79 have been used. These flames have been stabilized on a burner at a pressure of 6.7 kPa using argon as dilutant. A new mechanism for the oxidation of allene, propyne, 1,3-butadiene and cyclopentene has been proposed including the formation and decomposition of benzene and toluene. The main reaction pathways of formation of aromatics have been derived from flow rate analyses and have been compared for the three types of additives.

## Introduction

While the formation of PAHs and soot represents an important area of interest for kineticists for two decades, questions still remain even concerning the formation and the oxidation of the first aromatic compounds. In previous studies, the formation of benzene is mostly related to the reactions of $C_2$ (acetylene), $C_3$ or $C_4$ unsaturated species [1-5]. Nevertheless, some papers suggest a link between $C_6$ and $C_5$ cyclic species [6,7]. In a recent paper [8], we have investigated the reactions of allene and propyne, as they are precursors of propargyl radicals, which have an important role in forming benzene. In a second part [9], we have analyzed the reaction of 1,3-butadiene in a methane flame and putted in evidence an important production of benzene due to reactions of $C_4$ compounds under these particular conditions. To finish this work, we have studied the reactions of cyclopentene, which is a source of $C_5$ radicals [10]. The purpose of the present paper is to compare the results obtained in these three studies for the formation of benzene and toluene and their ways of formation. It is worth noting that the use of methane as the background and, consequently, of a flame rich in methyl radicals favors reactions which can be less important in other combustion systems.

## Experimental procedure and main results

The experiments were performed using an apparatus developed in our laboratory to study temperature and stable species profiles in a laminar premixed flat flame at low pressure [8]. This horizontal burner was housed in a water-cooled vacuum chamber evacuated by two primary pumps and maintained at 6.7 kPa by a regulation valve. This chamber was equipped of four quartz windows for an optical access, a microprobe for samples taking and a thermocouple for temperature measurements. The burner could be vertically translated, while the housing and its equipments were kept fixed.

Gas flow rates were regulated by RDM 280 Alphagaz and Bronkhorst (El-Flow) mass flow regulators. The $C_3$ and $C_4$ reactants and methane were supplied by Alphagaz - L'Air Liquide. Oxygen and argon were supplied by Messer. Liquid cyclopentene (provided by Fluka) flow rate was controlled by using a liquid mass flow controller, mixed to the carrier gas and then evaporated by passing through a single pass heat exchanger, the temperature of which was set above the boiling point of the mixture [10].

Temperature profiles were obtained using a PtRh (6%)-PtRh (30%) type B thermocouple (diameter 100 $\mu$m) coated with an inert layer to prevent catalytic effects. Radiative heat losses were corrected using the electric compensation method. The sampling probe was in silica with a hole of about 50 µm diameter. For temperature measurements in the flames perturbed by the probe, the distance between the junction of the thermocouple and the end of the probe was taken to about 100 µm. In the studied flames, the temperature ranged from 600 K close to the burner up to 2150 K.

Gas samples were collected in a Pyrex loop and directly obtained by connecting through a heated line the quartz probe to a volume, which was previously evacuated. The pressure drop between the flame and the inlet of the probe ensured reactions to be frozen. Stable species profiles were determined by gas chromatography. Chromatographs with a Carbosphere packed column and helium or argon as carrier gas were used to analyse $O_2$, $H_2$, CO and $CO_2$ by thermal conductivity detection and $CH_4$, $C_2H_2$, $C_2H_4$, $C_2H_6$ by flame ionisation detection (FID). Heavier hydrocarbons from $C_3$ to $C_7$ were analysed on a Haysep packed column by FID and nitrogen as gas carrier gas. The identification of these compounds was performed using GC/MS and by comparison of retention times when injecting the product alone in gas phase. Water and small oxygenated compounds, such as acetone, acetaldehyde and formaldehyde, were detected by GC-MS but not quantitatively analysed.

* Corresponding author: Pierre-Alexandre.glaude@ensic.inpl-nancy.fr
Associated Web site: http://www.ensic.inpl-nancy.fr/DCPR/

Due to problems of stability, a slightly higher dilution (55.6% argon instead of 42.9%) has been used in the flame of cyclopentene compared to the 3 other flames.

In the pure methane flame and in all the doped ones, quantified products included carbon monoxide and dioxide, hydrogen, ethane, ethylene, acetylene, propyne (p-$C_3H_4$), allene (a-$C_3H_4$), propene ($C_3H_6$) and propane ($C_3H_8$). The presence of each of the four additives strongly promotes the formation of acetylene and unsaturated $C_3$ species. Compared to a methane flame of same equivalence ratio, the maximum of acetylene mole fraction is almost multiplied by a factor 1.6 with the addition of allene, 2 with that of propyne and 1,3-butadiene and 2.5 with that of cyclopentene.

Several $C_{3+}$ products, which cannot be detected in the pure methane flame, have also been analysed:

- **in the case of $C_3$ additives**, butadienes (1,2-$C_4H_6$, 1,3-$C_4H_6$), 1-butene (1-$C_4H_8$), iso-butene (i-$C_4H_8$), 1-butyne (1-$C_4H_6$), vinylacetylene ($C_4H_4$) and benzene ($C_6H_6$) (toluene could be quantified for flames of richer mixtures, but was not detected in the flames studied here) [8],
- **in the case of 1,3-butadiene**, 1,2-butadiene, butynes (1-$C_4H_6$ and 2-$C_4H_6$), vinylacetylene, diacetylene ($C_4H_2$), 1,3-pentadiene (1,3-$C_5H_8$), 2-methyl-1,3-butadiene (isoprene, i$C_5H_8$), 1-pentene (1-$C_5H_{10}$), 3-methyl-1-butene (i$C_5H_{10}$), benzene and toluene ($C_7H_8$) [9],
- **in the case of cyclopentene**, butadienes, vinylacetylene, diacetylene, cyclopentadiene (c$C_5H_6$), 1,3-pentadiene, benzene and toluene [10].

## Description of the detailed kinetic model

The mechanism that was proposed to model the oxidation of allene, propyne, 1-3-butadiene and cyclopentene includes an up dated version of the mechanism that was built to model the oxidation of $C_3$-$C_4$ unsaturated hydrocarbons [8, 9, 11, 12], our previous mechanisms of the oxidation of benzene [13] and toluene [14] and a new mechanism for the oxidation of cyclopentene [10]. Thermochemical data were estimated using the software THERGAS developed in our laboratory [15], which is based on the additivity methods proposed by Benson [16].

*Reaction base for the oxidation of C3-C4 unsaturated hydrocarbons [8, 9]*

This $C_3$-$C_4$ reaction base, which was described in details in previous papers [8, 9], was built from a review of the recent literature and is an extension of our previous $C_0$-$C_2$ reaction base [17]. The $C_3$-$C_4$ reaction base includes reactions involving $C_3H_2$, $C_3H_3$, $C_3H_4$ (allene and propyne), $C_3H_5$, $C_3H_6$, $C_4H_2$, $C_4H_3$, $C_4H_4$, $C_4H_5$, $C_4H_6$ (1,3-butadiene, 1,2-butadiene, methyl-cyclopropene, 1-butyne and 2-butyne), $C_4H_7$ (6 isomers), as well as some reactions for linear and branched $C_5$ compounds and the formation of benzene. In this reactions base, pressure-dependent rate constants follow the formalism proposed by Troe [18] and efficiency coefficients have been included.

*Mechanisms for the oxidation of benzene and toluene*

Our mechanism for the oxidation of benzene contains 135 reactions and includes the reactions of benzene and of cyclohexadienyl, phenyl, phenylperoxy, phenoxy, hydroxyphenoxy, cyclopentadienyl, cyclopentadienoxy and hydroxycyclopentadienyl free radicals, as well as the reactions of ortho-benzoquinone, phenol, cyclopentadiene, cyclopentadienone and vinylketene, which are the primary products yielded [13].

The mechanism for the oxidation of toluene contains 193 reactions and includes the reactions of toluene and of benzyl, tolyl, peroxybenzyl (methylphenyl), alcoxybenzyl and cresoxy free radicals, as well as the reactions of benzaldehyde, benzyl hydroperoxyde, cresol, benzylalcohol, ethylbenzene, styrene and bibenzyl [14].

*New mechanism proposed for the oxidation of cyclopentene [10]*

We have considered the unimolecular reactions of cyclopentene, the additions of H-atoms and OH radicals to the double bonds and the H-abstractions by oxygen molecules and small radicals. Unimolecular reactions include dehydrogenation to give cyclopentadiene, decompositions by breaking of a C-H bond and isomerization to give 1,2-pentadiene. The reactions of cyclopentenyl radicals involved isomerizations, decompositions by breaking of a C-C bond to form linear $C_5$ radicals including two double bonds or a triple bond, the formation of cyclopentadiene by breaking of a C-H bond or by oxidation with oxygen molecules and terminations steps. Termination steps were written only for the resonance stabilized cyclopentenyl radicals: disproprtionnations with H-atoms and OH radicals gave cyclopentadiene, combinations with $HO_2$ radicals led to ethylene and $CH_2CHCO$ and OH radicals, and combinations with $CH_3$ radicals formed methylcyclopentene. The decomposition by breaking of a C-H bond of cyclopentyl radicals led to the formation of 1-penten-5-yl radicals. The isomerizations (for the radical stabilized isomer) and the decompositions by breaking of a C-C bond of the linear $C_5$ radicals were also written, while those by breaking of a C-H bond were not considered. The reactions of cyclopentadiene were part of the mechanism for the oxidation of benzene, but reactions for the consumption of methylcyclopentene and methylcyclopentadiene, obtained by recombination of cyclopentadienyl and methyl radicals had to be added.

*Ways of formation of aromatic compounds [8, 9, 10]*

In order to investigate the relative importance of the different channels, the formation of aromatic compounds was considered through the $C_3$, the $C_4$ and the $C_5$ pathways:

- **For the $C_3$ pathway**, we have used a value of $1.10^{12}$ $cm^3.mol^{-1}s^{-1}$ for the recombination of two propargyl radicals to give phenyl radicals and H-atoms, which is in good agreement with what has been recently proposed by Miller et al. [19] and Rasmussen et al. [20]. We have removed from our previous mechanism the addition of propargyl radicals to allene leading to benzene and H-atoms and the recombination between allene and propargyl radicals to produce phenyl radicals and two H-atoms, which were not considered by Rasmussen et al. [20] and which induced an overprediction of benzene in the case of the flame doped with allene.
- **For the $C_4$ pathway**, as proposed by Westmoreland et al. [2], all the reactions between $C_2$ species and n-$C_4H_3$, n-$C_4H_5$ radicals or 1,3-butadiene molecules and leading to aromatic and linear $C_6$ species have been considered. We have considered the dehydrogenation of 1,4-cyclohexadiene to give benzene with a rate constant proposed by Ellis and Freys [21], the H-abstractions from 1,4-cyclohexadiene by H-atoms and OH radicals to give $C_6H_7$ radicals with rate constants proposed by Dayma et al. [22] and the additions of H-atoms to benzyne ($C_6H_4$) to give phenyl radicals with a rate constant proposed by Wang et al. [5].
- **For the $C_5$ pathway**, as proposed by Lifshitz et al. [7], we have considered the isomerization between both radicals deriving from methylcyclopentadiene (cyclopentadienylmethylene and methylcyclo pentadienyl), and the formation of cyclohexadienyl radicals from cyclopentadienyl methylene radicals. We have not considered the formation of cyclohexadienyl radicals from the resonance stabilized methyl cyclopentadienyl radicals, as proposed by Marinov et al. [4], but their recombination with H-atoms was also added. In order to reproduce the important formation of toluene experimentally observed in the flame doped with cyclopentene, we have considered the formation of benzyl radicals from the addition of acetylene to cyclopendadienyl radicals with a rate constant about 3 times higher than what is usually considered for such an addition [23]. While the decomposition of benzyl radicals to form acetylene and cyclopendadienyl radicals, which was considered in our mechanism for the oxidation of toluene, has been studied by several authors [24-26] and has been shown to occur through a several steps mechanism, the reverse addition has never been directly investigated.

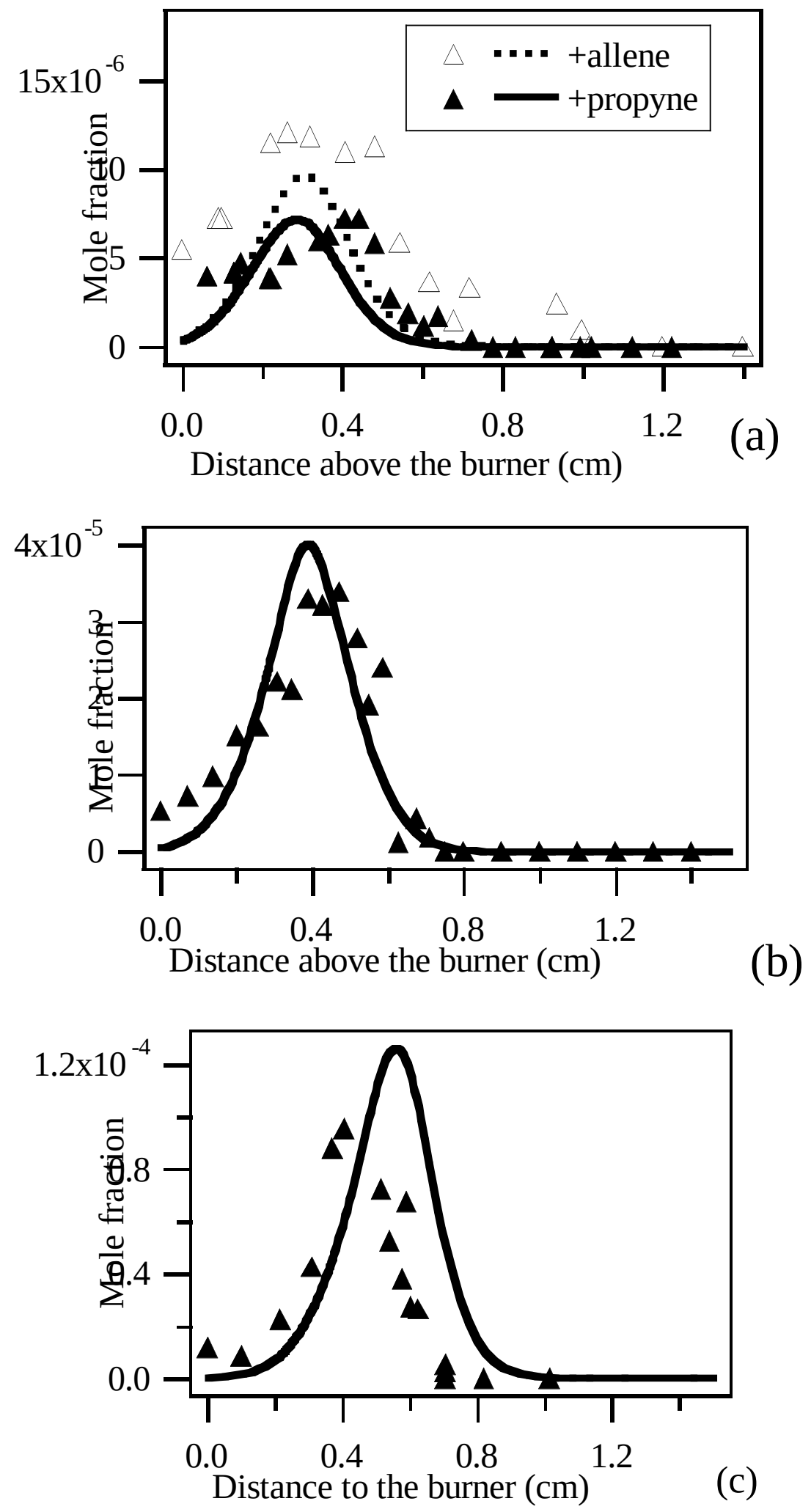

*Figure 1: Experimental and simulated profiles of benzene in a methane flame doped with (a) allene and propyne, (b) 1,3-butadiene and (c) cyclopentene. Symbols correspond to experiments, lines to simulations.*

**Comparison of the formation of benzene and toluene in the four flames**

All the simulations have been performed using PREMIX of CHEMKIN II [27] with estimated transport coefficients and the same kinetic mechanism. Comparisons between experimental results and simulations are displays for benzene in fig. 1 and for toluene in fig. 2. The experimental results show important changes in the formation of aromatic compounds depending on the additives. The flame showing the largest formation of benzene is that doped with cyclopentene. The maximum of the peak of benzene is three times smaller in the flame doped with 1,3-butadiene, ten times in the flame doped by allene and fifteen times in that doped by propyne.

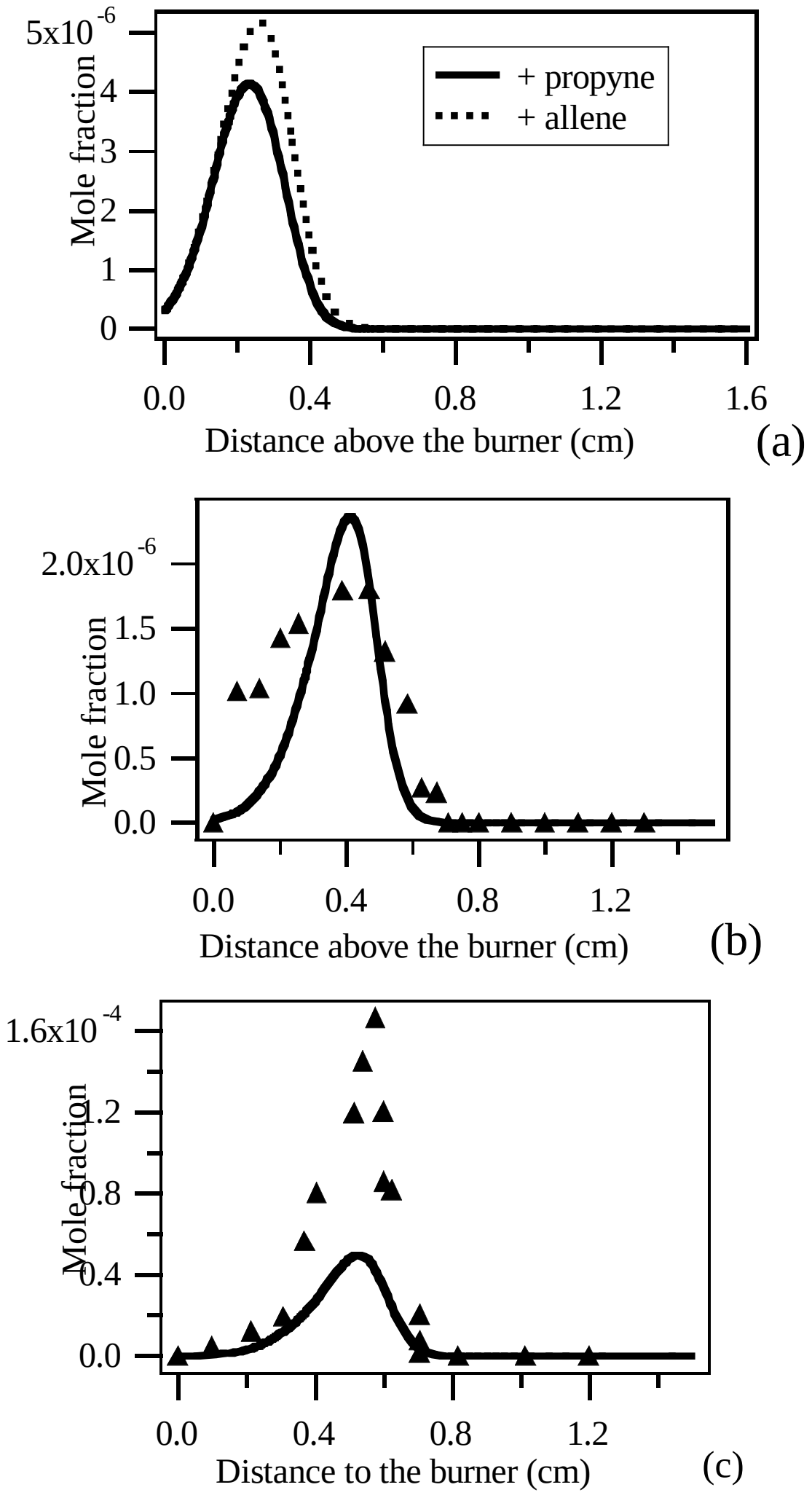

Figure 2: Experimental and simulated profiles of toluene in a methane flame doped with (a) allene and propyne, (b) 1,3-butadiene and (c) cyclopentene. Symbols correspond to experiments, lines to simulations.

The flame doped with 1,3-butadiene exhibits the lowest production of toluene. The maximum of the peak of toluene is two times higher in the flame doped with $C_3$ compounds and almost hundred times larger in that doped with cyclopentene.

The production of aromatic compounds occur the closest to the burner in the flames doped by $C_3$ compounds and the furthest in the flame seeded by cyclopentene.

Simulations reproduce correctly the formation of benzene and toluene except in the case of the flame doped with cyclopentene for which the production of toluene is undersestimated by a factor 3.

**Discussion**

Figure 3 presents a flow rate analysis for the production and consumption of aromatic compounds for the flames doped by allene (a similar picture is obtained for propyne), 1,3-butadiene and cyclopentene. This figure shows well that the ways of formation of benzene are very different depending on the additive:

- **in the allene flame**, benzene is produced by combination of H-atoms with phenyl radicals, which are obtained by self-combination of propargyl radicals,
- **in the 1,3-butadiene flame**, benzene is rapidly formed from the addition/cyclization of vinyl radicals to produce cyclohexadiene, which reacts either by dehydrogenation or by methatheses with H-atoms and OH radicals, followed by the decomposition of the obtained cyclic $C_6H_7$ radicals,
- **in the cyclopentene flame**, the major source of benzene is the isomerization of cyclopentadienyl methylene radicals, which are obtained either directly by H-abstractions by H-atoms or OH radicals from methylcyclopentadiene or by isomerisation from the resonance stabilized methyl cyclopentadienyl radicals, which are also obtained by H-abstractions by H-atoms or OH radicals from methylcyclopentadiene.

A particular interest of the flames doped by 1,3-butadiene and cyclopentene is that the production of benzene is mainly due to other reactions than the $C_3$ pathway as it is usually the case in the flames of the literature [20].

Differences are also encountered for the ways of formation of toluene:

- **in the allene flame**, toluene is mainly produced from the recombination of methyl and phenyl radicals,
- **in the 1,3-butadiene and cyclopentene flames**, the most important way to give toluene is the addition of resonance stabilized cyclopentadienyl radicals to acetylene.

In the $C_3$ flames, benzene and toluene are primary products deriving directly from propargyl and phenyl radicals. That explains why their formation occurs early in the flame. In these flames, the production of benzene is relatively weak because the self recombination of propargyl radicals, at the low pressure studied here, leads mainly to phenyl radicals, which react rapidly with oxygen molecules. It is worth noting that our simulation using a value of the rate constant for the self-combination of propargyl radicals close to that now admitted in the literature leads to a good agreement for the formation of benzene.

The rate constant of the recombination of methyl and phenyl radicals, which is not well known [14] is a sensitive parameters to predict the formation of toluene. Many similarities exist between allene and propyne flames in both reactivity and products formation. Nevertheless, the pool of small radicals is slightly larger in the propyne flame, which leads to a faster consumption of propargyl radicals and a lower formation of benzene than in the flame seeded by allene [8].

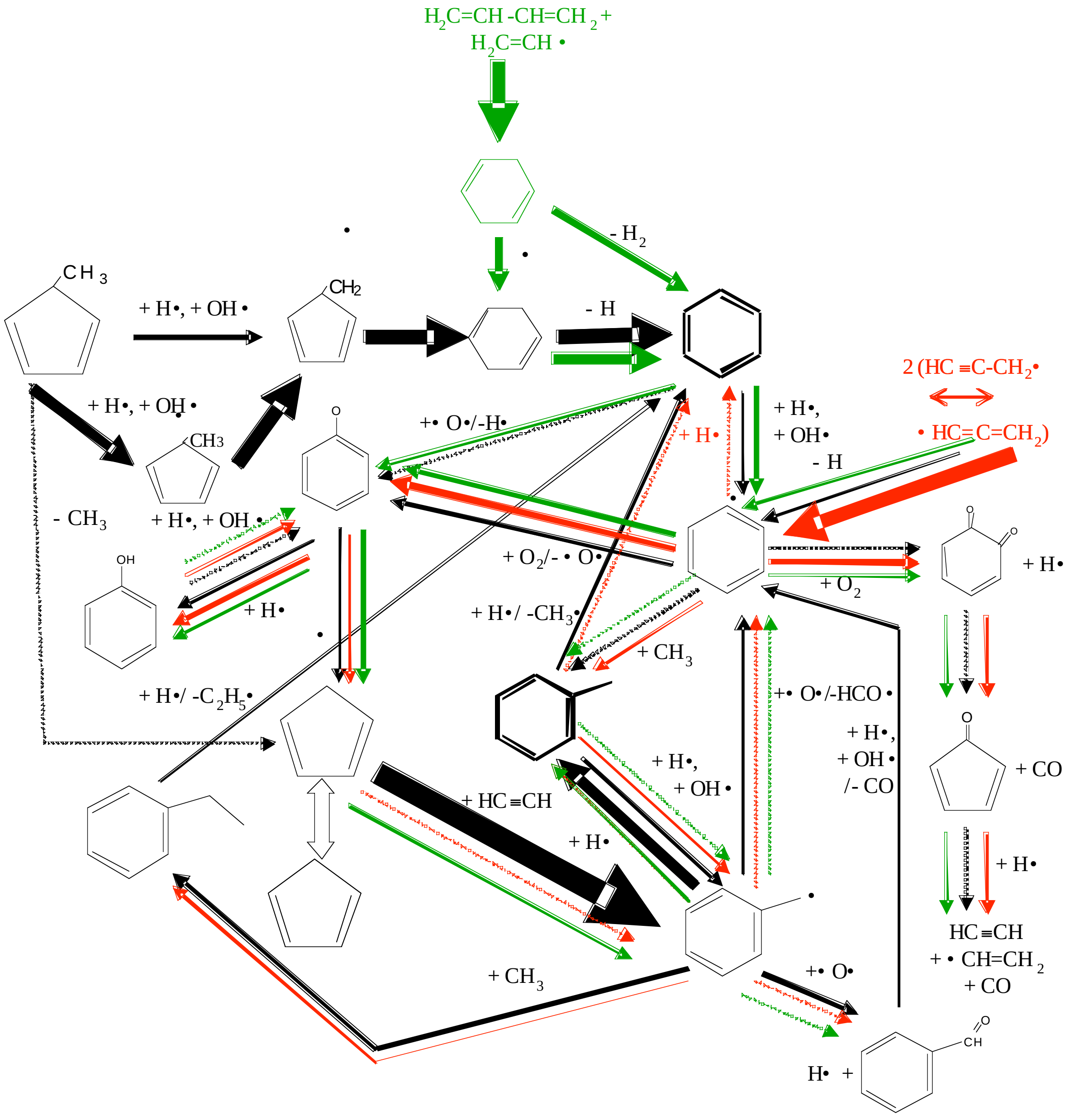

*Figure 3: Flow rate analysis for the formation and consumption of benzene and toluene for a position in the flame corresponding to the peak of benzene profile. The size of the arrows is proportional to the relative flow rates. Red arrows are related to the flame doped with allene (T= 1236 K) , green arrows to that doped with 1,3-butadiene (T= 1483 K) and black arrows to that seeded with cyclopentene (T= 1354 K, the lower temperature compared to the case of 1,3-butadiene is due to a higher dilution).*

Cyclopendadienyl radicals are rapidly obtained from the decomposition of phenoxy radicals, which are produced by the reaction of phenyl radicals with oxygen molecules. Due to their resonance stabilization, they are present with an important concentration in every system containing phenyl radicals and benzene. The formation of toluene by addition of cyclopentadienyl radicals to acetylene could then be of importance in many combustion systems rich in acetylene.

When the concentration of toluene is high enough, a non negligible formation of benzene occurs also through the ipso-addition of H-atoms with elimination of methyl radicals. The concentrations of benzene and toluene are thus closely linked.

Figure 3 shows that the addition of allene, 1,3-butadiene and cyclopentene leads also to the formation of benzoquinone, phenol, cyclopentadienone and benzaldehyde, which have not been analysed experimentally here.

## Conclusion

This paper compares the formation of benzene and toluene in a rich (Φ=1.8) methane flame seeded with allene, propyne, 1,3-butadiene and cyclopentene, respectively, and shows that the amount of aromatic compounds formed are very different depending on the additive. This effect is due to differences in the pathways of formation of these aromatic species depending on the dopant. For the flames doped with $C_4$ and $C_5$ compounds, the self-combination of propargyl is not the preponderant channel of formation of benzene.